\documentclass[sigconf,nonacm]{style/acmart} 

\AtBeginDocument{%
  }
    
\begin{document}

\title{DS@GT at TREC TOT 2025: Bridging Vague Recollection with Fusion Retrieval and Learned Reranking}
\newcommand{\affiliationgatech}{%
  \affiliation{%
    \institution{Georgia Institute of Technology}
    \city{Atlanta}
    \state{Georgia}
    \country{USA}}%
}

\author{Wenxin Zhou}
\affiliationgatech
\email{wzhou77@gatech.edu}
\orcid{0009-0002-3325-3357}

\author{Ritesh Mehta}
\affiliationgatech
\email{rmehta307@gatech.edu}

\author{Anthony Miyaguchi}
\affiliationgatech
\email{acmiyaguchi@gatech.edu}
\orcid{0000-0002-9165-8718}

\begin{abstract}
  We develop a two-stage retrieval system that combines multiple complementary retrieval methods with a learned reranker and LLM-based reranking, to address the TREC Tip-of-the-Tongue (ToT) task. In the first stage, we employ hybrid retrieval that merges LLM-based retrieval, sparse (BM25), and dense (BGE-M3) retrieval methods. We also introduce topic-aware multi-index dense retrieval that partitions the Wikipedia corpus into 24 topical domains. In the second stage, we evaluate both a trained LambdaMART reranker and LLM-based reranking. To support model training, we generate 5000 synthetic ToT queries using LLMs. Our best system achieves recall of 0.66 and NDCG@1000 of 0.41 on the test set by combining hybrid retrieval with Gemini-2.5-flash reranking, demonstrating the effectiveness of fusion retrieval.
\end{abstract}


\maketitle

\section{Introduction}

The TREC Tip-of-the-Tongue (ToT) track addresses the known-item search problem where a user has previously encountered an item, such as a movie, landmark, or celebrity, but cannot recall a specific identifier like a title or name to easily find it. 
The user's query is often a verbose, natural language description of the item they are trying to find. 
These descriptions are complex, sometimes containing multi-hop reasoning, expressions of uncertainty, and even false memories, which makes it difficult for traditional keyword-based search systems to handle effectively. 
The primary task of the ToT track is to develop systems that can successfully identify the "known-item" from these elaborate and often imprecise descriptions across various domains.

\begin{figure}[ht]
    \centering
    \includegraphics[width=0.9\linewidth]{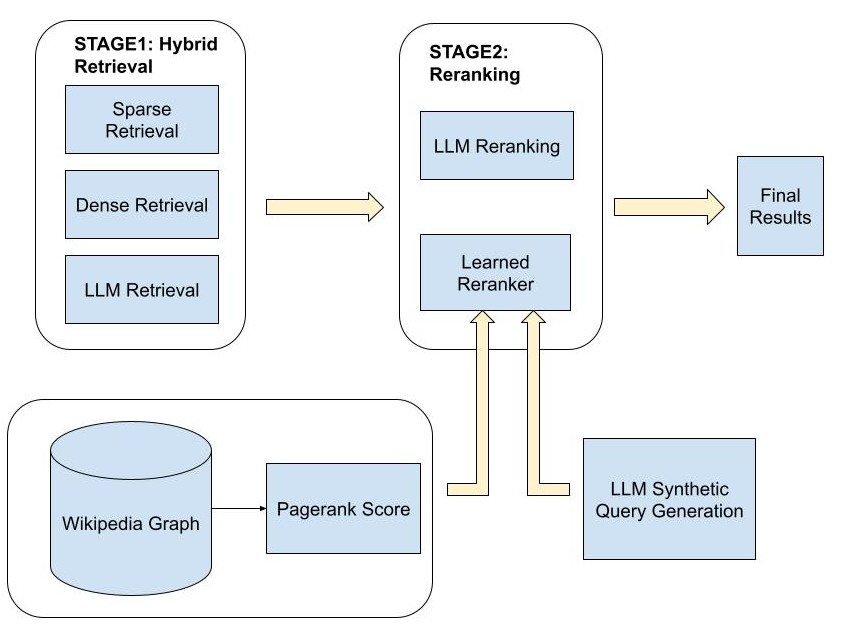}
    \caption{The two-stage retrieval architecture for ToT search.}
    \label{fig:two_stage_architecture}
\end{figure}

In our paper, we address the ToT challenge using two-stage retrieval, as shown in Figure~\ref{fig:two_stage_architecture}.
In the first stage, we use a hybrid retrieval method that combines LLM-based, sparse, and dense retrieval results. We explore topic-aware multi-index dense retrieval that leverages semantic document organization to improve search efficiency. For this method, the Wikipedia corpus is partitioned into 24 topical domains, enabling query-specific routing to relevant document subsets.

In the second stage, we experiment with an LLM-based reranker and a learned LambdaMART reranker. The learned reranker leverages various features such as dense and sparse retrieval scores, page view count, and network-theoretic measures (i.e., PageRank).
For data augmentation, we use LLMs to generate 5000 synthetic open-domain ToT queries to train the learned reranker.

Our findings reveal that the hybrid retrieval approach significantly enhances recall compared to individual retrieval methods. The LLM reranker achieves the best overall performance. The LambdaMART reranker shows notable improvements over baseline methods in terms of recall. The topic-aware routing approach demonstrates a trade-off between efficiency and effectiveness, achieving competitive recall with a reduced search space. The implementation can be found at \url{https://github.com/dsgt-arc/trec-tot-2025}. 

\section{Methodology}

\subsection{Baseline Methods}

For sparse retrieval, we use the PyTerrier BM25 baseline results provided by the ToT track organizers.

For dense retrieval, we use the BGE-M3 model \cite{bge-m3} to generate query embeddings, and perform dot product similarity search on pre-computed Wikipedia paragraph embeddings from the June 2024 Wikipedia dump \cite{upstash-wikipedia}.
We aggregate paragraph-level scores to article-level scores by taking the maximum paragraph score for each article (i.e max pooling).
Although the corpus timestamp differs from the official ToT corpus dated 2023-11-01, we believe the variation is minimal and unlikely to significantly impact overall performance.

\subsection{Network Construction}

We construct a composite directed graph $G=(V, E)$ over Wikipedia page links by creating a weighted superposition of three topological signals.
The edge weight $w(u,v)$ between two article nodes $u$ and $v$ is defined as the sum of three components:

\begin{itemize}
    \item \textbf{Direct Citations:} Links where article $u$ directly cites article $v$, denoted as $(u) \to (v)$.
    \item \textbf{Meta-Path Connections:} Indirect links formed via a non-article intermediate node $m$ (e.g., a Category or Redirect), denoted as the path $(u) \to (m) \to (v)$.
    \item \textbf{Semantic Nearest Neighbors:} Edges connected to the $k=15$ nearest neighbors based on cosine similarity of BGE-M3 embeddings.
\end{itemize}

Formally, the composite adjacency matrix $W$ is calculated as $W = \alpha_1 A_{direct} + \alpha_2 A_{meta} + \alpha_3 A_{sem}$, where $A$ represents the adjacency matrix for each respective signal and $\alpha_1=1.0$, $\alpha_2=0.5$, and $\alpha_3=0.25$.

\begin{table}[h]
    \centering
    \caption{Summary statistics for graph centrality measures ($N=6.6M$). The distribution is highly skewed, with the mean substantially larger than the median.}
    \label{tab:graph-stats}
    \resizebox{\columnwidth}{!}{%
    \begin{tabular}{lrrrr}
        \toprule
        \textbf{Metric} & \textbf{Mean} & \textbf{Std. Dev.} & \textbf{Min} & \textbf{Max} \\
        \midrule
        PageRank & $1.5\text{e-}7$ & $1.4\text{e-}5$ & $2.1\text{e-}8$ & $9.6\text{e-}3$ \\
        In-Degree & 161.9 & 8,760.7 & 15 & 4,294,807 \\
        Out-Degree & 156.7 & 215.3 & 0 & 10,555 \\
        \bottomrule
    \end{tabular}%
    }
\end{table}

PageRank is a measure of importance within a network derived from the link structure such that nodes that are linked to by nodes that are frequently visited are ranked higher \cite{page1999pagerank}.
The PageRank vector $\vec{\pi}$ is computed by finding the principal eigenvector of the stochastic matrix $P$ such that $\vec{\pi}P = \vec{\pi}$.
The matrix $P$ is constructed by row-normalizing the composite adjacency matrix $W$ with a uniform teleportation matrix $T$, such that:

\begin{equation}
P_{ij} = 
\begin{cases} 
\alpha \frac{W_{ij}}{\sum_k W_{ik}} + (1-\alpha)T_{ij} & \text{if } \sum_k W_{ik} > 0 \\
T_{ij} & \text{if } \sum_k W_{ik} = 0 
\end{cases}
\quad \text{where } T_{ij} = \frac{1}{N}
\end{equation}

We compute PageRank on this weighted graph using a damping factor of $0.85$ via rustworkx (200 iterations, tolerance $1 \times 10^{-10}$) and use it as a feature for reranking \cite{treinish2021rustworkx}.

\subsection{Data Augmentation}

Based on the ToT query elicitation work from He et al. \cite{tot-elicitation-sigir2025}, we use LLMs to generate synthetic tip-of-the-tongue queries to augment the training dataset. We revised the prompt (see Appendix~\ref{sec:prompt}) to generate open-domain queries. The prompt was adapted in two aspects. First, it removes domain-specific keywords such as "movie" and "landmark" and uses general terms such as "object" and "name." Second, it asks the LLM to return the query inside a code block, which helps extract the query from the LLM response. The LLM models used are gemini-2.5-flash-lite \cite{google2025gemini}, o4-mini \cite{openai2025o4mini} and gpt-4o \cite{gpt4o2025}.

Next, we sample approximately 5,000 entities from the corpus. We use Wikipedia article pageview and infobox template information to guide the sampling process to ensure diversity and popularity of the selected entities.

We aggregate user pageview counts from October 2022 to October 2023. To ensure relevance, we retain only pages that rank in the top 20\% by total views. Additionally, we filter out stub and short articles by keeping only those with word counts exceeding the 25th percentile. We also exclude articles with more than 5,000 words to avoid extremely long pages.
We only use pages that have infobox template information, which is inspired by He et al. \cite{tot-elicitation-sigir2025}. We organized entities into infobox template categories and applied the following sampling strategy for each category:
\begin{itemize}
  \item For categories with more than 1000 entities, select 5 entities.
  \item For categories with 500-1000 entities, select 4 entities.
  \item For categories with 100-500 entities, select 3 entities.
  \item For categories with 10-100 entities, select 1-2 entities.
\end{itemize}

This process selects 5,000 entities, and we use the gemini-2.5-flash-lite model to generate synthetic queries for them. This augmented dataset is then split into train, dev, and test sets with 75\%, 15\%, and 10\%, respectively.

\subsection{First-Stage Retrieval}

\subsubsection{LLM Retrieval}

We use LLMs as one of the methods for first-stage retrieval. In the prompt, we ask the LLM to return a single  or up to 20 Wikipedia article titles that match the tip-of-the-tongue query, with relevance scores from 1-5.
We request the LLM to return results in JSON format to make extracting article names from the response easier. The full prompt can be found in Appendix~\ref{sec:prompt_retrieval}.

Next, we look up the Wikipedia title in the corpus using code adapted from the 2024 ToT baseline solution. First, we check for exact title name matches. Then, we perform inexact name matching, which matches using redirect or alias names. Finally, if both fail, we remove the parenthetical portion of the title and perform exact and inexact name matching with the shortened title.

\subsubsection{Topic-Aware Multi-Index Dense Retrieval}
\begin{figure}[ht]
    \centering
    \includegraphics[width=0.9\linewidth]{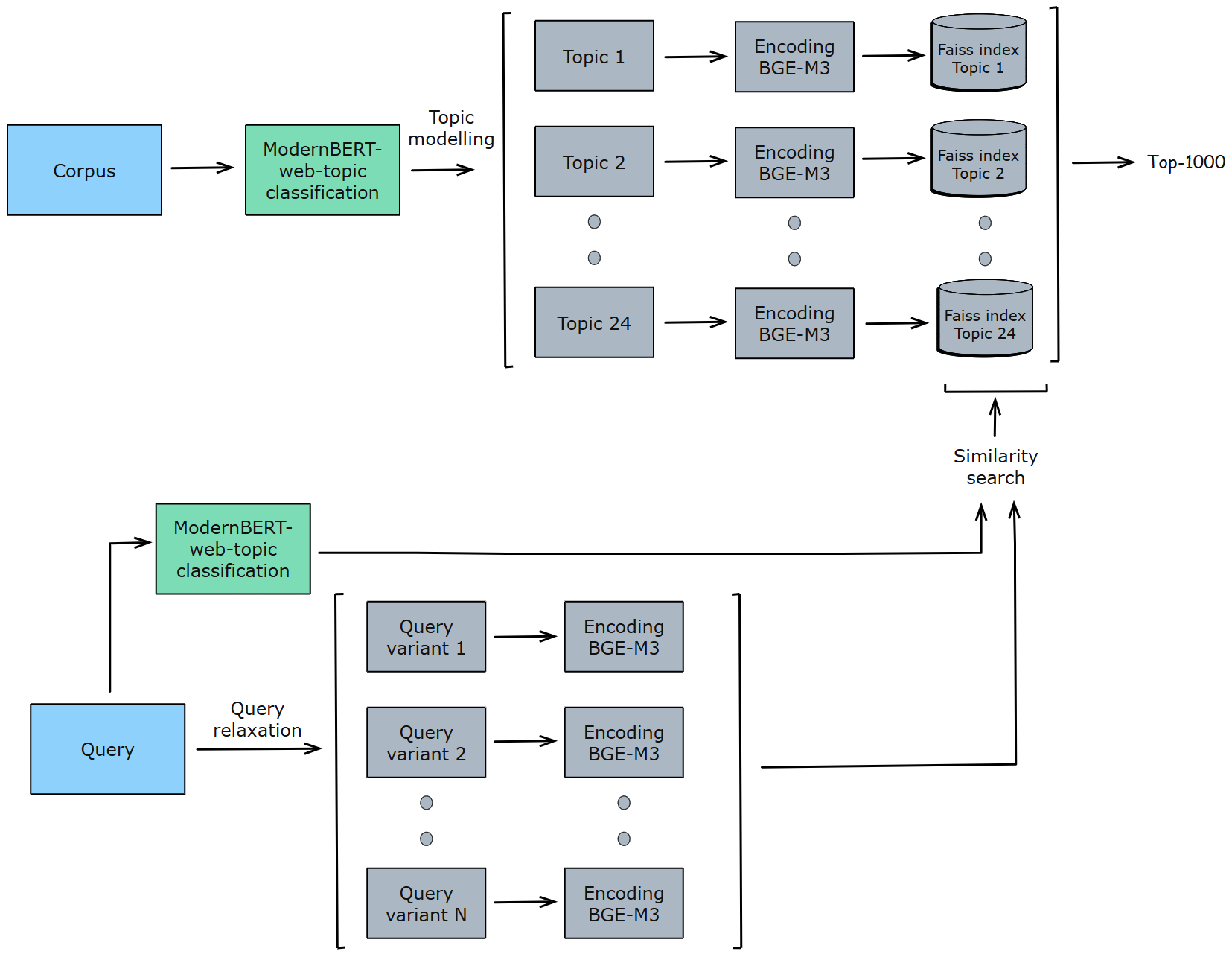}
    \caption{Topic-aware multi-index pipeline.}
    \label{fig:topic_aware_architecture}
\end{figure}

In this approach, we organized the Wikipedia corpus into 24 topical domains using ModernBERT-web-topics classification, with quality filtering to remove list articles, disambiguation pages, and content lacking substantive explanatory content.
For each topic, we construct separate FAISS indexes using BGE-M3 embeddings (1024-dimensional), processing long articles with overlapping chunking and averaging chunk embeddings to produce L2-normalized vectors for efficient cosine similarity search.
To handle the incomplete and uncertain nature of tip-of-the-tongue queries, we generate multiple relaxed query variants using Llama 3.1-8B-Instruct with carefully designed prompts that generalize queries while preserving core semantics. 
For queries where Llama produced unsatisfactory results, we used Gemini's chat interface to manually generate high-quality variants following the same generalization principles. 
This complete pipeline is illustrated in Figure~\ref{fig:topic_aware_architecture}.

The retrieval pipeline classifies each query variant to one of the 24 topics using ModernBERT-web-topics, then encodes it with BGE-M3 and searches only the corresponding topic-specific FAISS index for the top-1000 results per variant.
Results across all variants are aggregated and deduplicated by document ID, retaining the highest similarity score for each unique document.
The final top-1000 documents are selected and formatted in TREC format.
This topic-specific routing reduces the search space by focusing on relevant topical subsets of the corpus, while leveraging GPU acceleration for efficient batch processing of embeddings and index operations.

\subsubsection{Hybrid Retrieval}
We merge the sparse, dense and LLM retrieval results using a round-robin strategy. Specifically, we iterate through the ranked lists from each retrieval method and alternate which method contributes the next candidate to the merged result. For example, we take the top-ranked result from LLM retrieval, followed by the top-ranked result from dense retrieval, then the top-ranked result from sparse retrieval, and repeat this cycle. When a method runs out of candidates, we continue cycling through the remaining methods. When we encounter duplicate candidates, we keep the one with the highest rank from its original list.

\subsection{Reranking}

\subsubsection{LLM Reranker}
We use an LLM to perform listwise reranking. The implementation is based on the RankLLM library\cite{sharifymoghaddam2025rankllm}, where a sliding window algorithm is used to split a large set of documents into groups so that the context can fit within the LLM's context window.
We tested various context settings, such as title only, first paragraph, first sentence, and full (first 1500 characters), and adjusted the sliding window size and stride accordingly.
We also experimented with various models such as Gemma and Gemini.

\subsubsection{LambdaMART Reranker}

We trained a LambdaMART reranker \cite{wu2008ranking} using the xgboost library \cite{chen2016xgboost}, and explored multiple sampling strategies and feature sets for training. The final LambdaMART model uses five features: dense retrieval score (from the BGE-M3 model), PyTerrier BM25 score for each document-query pair, normalized pageview count and pagerank score for the document, and the query's word count. 

For sampling, we adopted a pseudo-relevance approach. Each query's golden document is assigned a score of 2. Five pseudo-relevant documents are randomly selected from the top-10 results of either dense or sparse retrieval and assigned a score of 1. Ten irrelevant documents are randomly sampled from ranks 10 to 100 in the retrieval results and given a score of 0. This ensures that each query in the training set is associated with one golden document, five pseudo-relevant documents, and ten irrelevant documents, providing a balanced distribution to mitigate class imbalance and reduce overfitting.

The training and validation data consist of official train set (143 queries) and the first 200 queries of dev3 set, and the training set of our generated queries (around 4000 queries). Among those, 80\% are used for training and 20\% for validation.

\section{Results}

\subsection{Synthetic Query Correlation Study}
\begin{figure}[ht]
    \centering
    \includegraphics[width=0.9\linewidth]{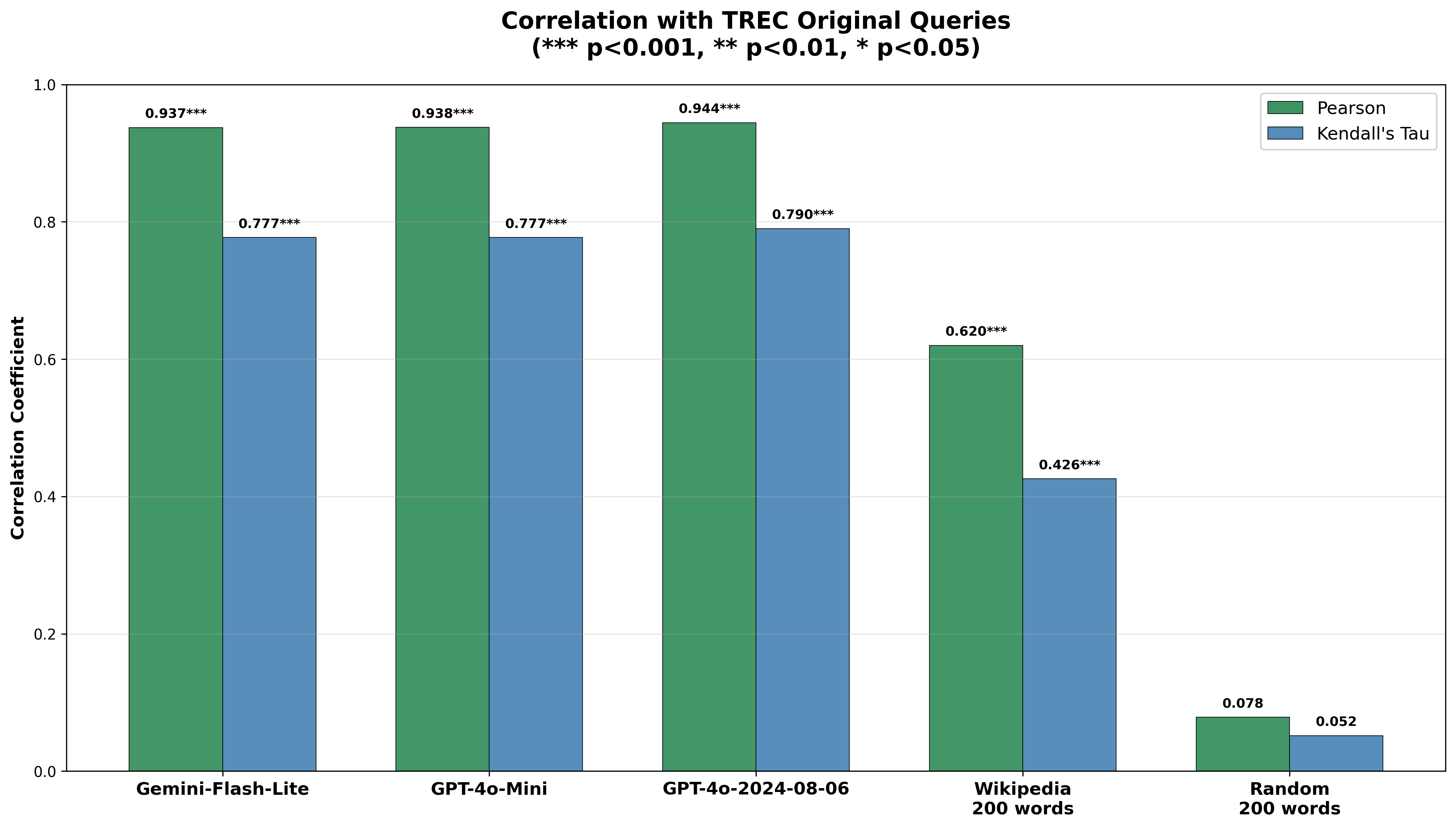}
    \caption{Pearson and Kendall's Tau correlation for synthetic and original queries using \textit{all-MiniLM-L6-v2} embeddings.}
    \label{fig:synthetic_query_correlation}
\end{figure}

To validate that our adapted LLM query generation method produces queries that are similar to the official dataset, we test our method on the Wikipedia articles associated with the first 100 queries of the dev3 set. We also use two baselines for comparison. One baseline randomly generates 200 words as the query, and the other uses the first 200 words of the Wikipedia article as the query. The correlation metrics, the Pearson and Kendall's Tau scores are calculated on the embeddings of the original queries and the generated queries, as shown in Figure~\ref{fig:synthetic_query_correlation}. \textit{all-MiniLM-L6-v2} is used as the embedding model.

All LLM-generated queries (Gemini and GPT-4o variants) show very high correlation with original queries (Pearson $>$ 0.93 and Tau $>$ 0.77), suggesting they capture similar semantic content and query characteristics. In comparison, Wikipedia text (the first 200 words of the Wikipedia article) shows moderate correlation. This result demonstrates that the queries generated with the new method have strong correlation with the original queries.

\begin{figure}[ht]
    \centering
    \includegraphics[width=0.9\linewidth]{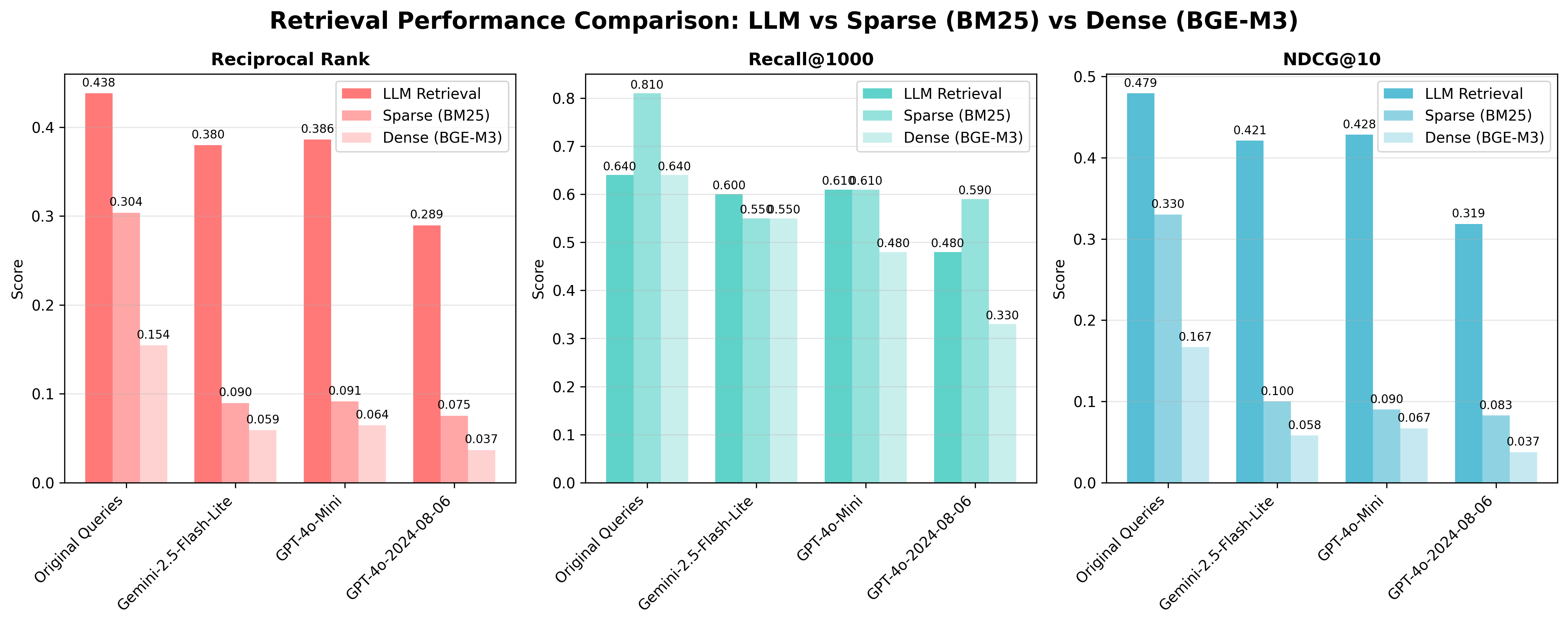}
    \caption{Retrieval performance comparison for synthetic and original queries using various retrieval methods.}
    \label{fig:synthetic_query_retrieval}
\end{figure}

We further compare the retrieval results of the original and generated queries using sparse, dense, and LLM retrieval methods (as shown in Figure~\ref{fig:synthetic_query_retrieval}). The results show that the LLM-generated queries achieve comparable retrieval performance to the original queries across all retrieval methods, with one outlier: the recall@1000 for dense retrieval on the original dataset is significantly better than that of the LLM synthetic queries. This further validates the effectiveness of our adapted LLM query generation method.

\subsection{LLM Retrieval Results}
We experiment with several LLMs and compare redirect and alias matching methods. Table~\ref{tab:llm_retrieval_results} shows the results for Gemini-2.5-flash and GPT-OSS models with the two title matching methods across various datasets.

\begin{table}[ht]
\centering
\small
\caption{LLM Retrieval Results: Redirect vs Alias Matching}
\begin{tabular}{lcccc}
\toprule
Dataset & Model & Method & R@1000 & NDCG@10 \\
\midrule
dev1 & Gemini & Redirect & 0.1408 & 0.1143 \\
dev1 & Gemini & Alias & 0.1620 & 0.1064 \\
dev1 & GPT-OSS & Redirect & 0.0634 & 0.0508 \\
dev1 & GPT-OSS & Alias & 0.0704 & 0.0406 \\
\midrule
dev2 & Gemini & Redirect & 0.1678 & 0.1437 \\
dev2 & Gemini & Alias & 0.1748 & 0.1266 \\
dev2 & GPT-OSS & Redirect & 0.0699 & 0.0526 \\
dev2 & GPT-OSS & Alias & 0.0909 & 0.0398 \\
\midrule
dev3 & Gemini & Redirect & 0.6343 & 0.5887 \\
dev3 & Gemini & Alias & 0.6437 & 0.5168 \\
dev3 & GPT-OSS & Redirect & 0.4944 & 0.4346 \\
dev3 & GPT-OSS & Alias & 0.4869 & 0.3765 \\
\bottomrule
\end{tabular}
\label{tab:llm_retrieval_results}
\end{table}

\subsection{Hybrid Retrieval Results}

Our hybrid retrieval approach demonstrates significant improvements over individual retrieval methods across all datasets by combining LLM, sparse (BM25), and dense (BGE-M3) retrieval using round-robin merging.

\begin{table}[ht]
\centering
\small
\caption{Hybrid Retrieval Performance at Recall@1000}
\begin{tabular}{lccccc}
\toprule
Dataset & LLM & BM25 & Dense & Hybrid & Improvement \\
\midrule
Dev1 & 0.1429 & 0.4507 & 0.4789 & 0.5000 & 4.4\% \\
Dev2 & 0.1727 & 0.4545 & 0.5105 & 0.5664 & 11.0\% \\
Dev3 & 0.6355 & 0.7705 & 0.6474 & 0.8302 & 7.8\% \\
\bottomrule
\end{tabular}
\label{tab:hybrid_retrieval_results}
\end{table}

\begin{figure}[ht]
\centering
\includegraphics[width=0.9\linewidth]{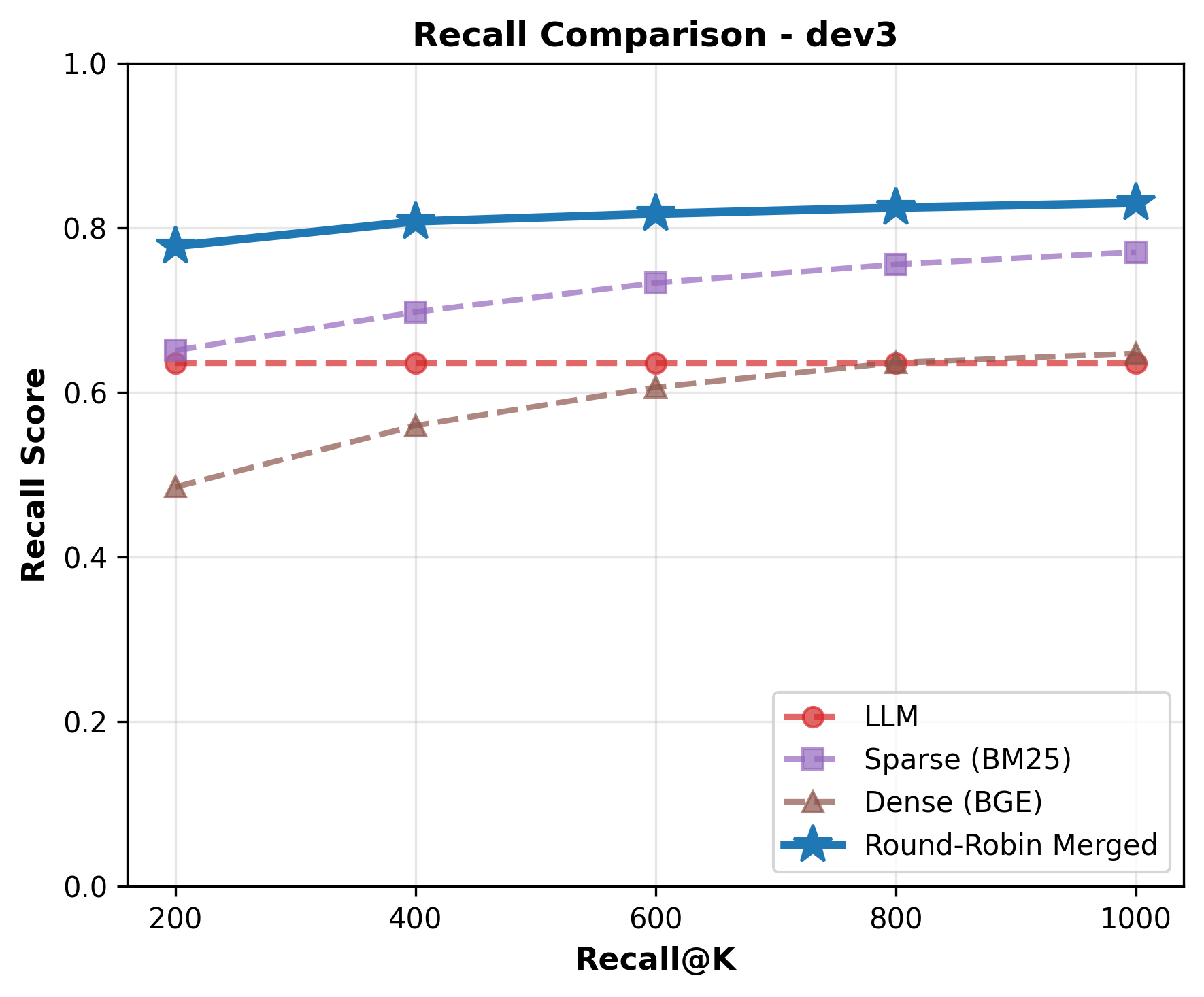}
\caption{Recall comparison of individual and hybrid retrieval methods on dev3 set.}
\label{fig:dev3_merged_recall}
\end{figure}

Table~\ref{tab:hybrid_retrieval_results} summarizes recall@1000 across all datasets. The hybrid approach improves over the best individual method on each dataset. Dev3 shows the strongest absolute performance, while Dev2 demonstrates the largest relative improvement at 11\%. Figure~\ref{fig:dev3_merged_recall} illustrates the recall comparison at various cutoffs for the dev3 set.

These results show that the round-robin merging strategy effectively combines the complementary strengths of different retrieval methods, providing notable recall improvements.

\subsection{LLM Reranking Results}

We evaluate LLM-based listwise reranking on the first 100 queries of the dev3 set using full document context with sliding window size 20 and stride 10. Table~\ref{tab:llm_reranking_results} compares the reranking performance of various models on sparse (PyTerrier BM25) and dense (BGE-M3) retrieval baselines with recall@1000 of 0.81 and 0.64, respectively.

\begin{table}[ht]
\centering
\small
\caption{LLM Reranking Results on dev3 Set (first 100 queries)}
\begin{tabular}{lcccc}
\toprule
Baseline & Reranker Model & RR & NDCG@1000 \\
\midrule
Sparse & Gemma-3-27B-QAT & 0.5104 & 0.5534 \\
Sparse & Gemma-3-12B-QAT & 0.4832 & 0.5307 \\
Sparse & Mistral-Nemo-Instruct-2407 & 0.4485 & 0.5018 \\
Sparse & Llama-3.1-8B-Instruct & 0.4411 & 0.4911 \\
Sparse & Gemma-3-4B-QAT & 0.3098 & 0.3820 \\
Sparse & None (baseline) & 0.3038 & 0.3894 \\
Sparse & Gemma-3-1B-QAT & 0.0868 & 0.1759 \\
\midrule
Dense & Gemma-3-27B-QAT& 0.3393 & 0.3774 \\
Dense & Gemma-3-12B-QAT & 0.3318 & 0.3717 \\
Dense & Gemma-3-12B & 0.3227 & 0.3647 \\
Dense & None (baseline) & 0.1545 & 0.2315 \\
\bottomrule
\end{tabular}
\label{tab:llm_reranking_results}
\end{table}

Most LLM rerankers outperform the baseline in both retrieval settings. For sparse retrieval, Gemma-3-27B-QAT achieves an RR of 0.5104 (68\% improvement over the baseline of 0.3038). For dense retrieval, Gemma-3-27B-QAT achieves an RR of 0.3393 (120\% improvement over the baseline of 0.1545). Larger models consistently outperform smaller variants. All models with 8B+ parameters show strong performance.

\subsection{LambdaMART Reranker}

We train LambdaMART rerankers using xgboost with both grid search and random search for hyperparameter tuning. The training details can be found in Appendix~\ref{sec:lambdamart_tuning}. The models are trained with pairwise loss over 100 rounds using NDCG as the evaluation metric. We select the grid search variant (v5) as our final model, as it generalizes better to the unseen dev and synthetic datasets than the random search variant (v6).

\begin{figure}[ht]
    \centering
    \includegraphics[width=0.9\linewidth]{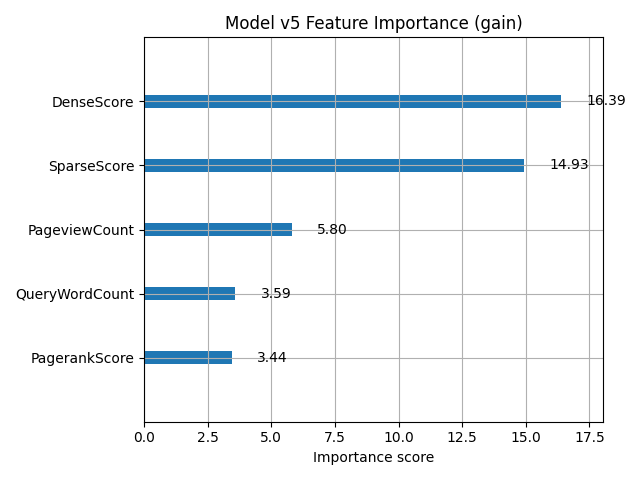}
    \caption{Feature importance for LambdaMART v5 model.}
    \label{fig:lambda_v5_importance}
\end{figure}

Table~\ref{tab:lambdamart_recall_results} presents recall improvements across all datasets on sparse (PyTerrier BM25) and dense (BGE-M3) baseline retrieval. Figure~\ref{fig:lambda_v5_importance} shows the feature importance weight of the v5 model.

\begin{table}[ht]
\centering
\caption{LambdaMART v5 Reranked Recall@10 and Recall@100. The arrow ($\to$) indicates improvement from baseline to reranked results. Synthetic refers to the dataset from our generated synthetic queries.}
\begin{tabular}{lccc}
\toprule
Dataset & Baseline & Recall@10 & Recall@100 \\
\midrule
Train & Sparse & 0.10 → 0.15 & 0.24 → 0.29 \\
Train & Dense & 0.08 → 0.12 & 0.20 → 0.29 \\
\midrule
Dev1 & Sparse & 0.11 → 0.14 & 0.22 → 0.25 \\
Dev1 & Dense & 0.10 → 0.14 & 0.23 → 0.27 \\
\midrule
Dev2 & Sparse & 0.15 → 0.15 & 0.23 → 0.26 \\
Dev2 & Dense & 0.11 → 0.16 & 0.27 → 0.27 \\
\midrule
Dev3 & Sparse & 0.43 → 0.49 & 0.60 → 0.66 \\
Dev3 & Dense & 0.21 → 0.43 & 0.40 → 0.58 \\
\midrule
Synthetic-dev & Sparse & 0.15 → 0.21 & 0.24 → 0.34 \\
Synthetic-dev & Dense & 0.12 → 0.20 & 0.27 → 0.38 \\
\bottomrule
\end{tabular}
\label{tab:lambdamart_recall_results}
\end{table}

Across all datasets, LambdaMART v5 consistently improves recall@10 and recall@100 over baseline BM25 and dense retrieval. For example, on the dev3 set, reranking increases recall@100 from 0.60 to 0.66 for sparse and from 0.40 to 0.58 for dense retrieval. Based on the feature importance gain for the LambdaMART v5 model, the dense and sparse retrieval scores are the most influential features (16.39 and 14.93, respectively), while pageview count (5.80) moderately contributes to reranking. This indicates that the learned reranker primarily relies on quality signals from the initial retrieval stage rather than document popularity.

\subsection{End-to-End System Results}

We submitted 7 systems to the TREC Tip-of-the-Tongue 2025 challenge. Table~\ref{tab:end_to_end_results} summarizes their performance on the official test set:

\begin{itemize}
  \item \textbf{bge-m3:} Dense retrieval baseline using BGE-M3 embeddings with cosine similarity search on Wikipedia passages from the June 2024 dump.
  
  \item \textbf{top\_model\_dense:} First-stage dense retrieval enhanced with topic-aware routing, partitioning the corpus into 24 topic-specific FAISS indexes with query classification and multi-variant query expansion.
  
  \item \textbf{gemini-retrieval:} LLM retrieval using Gemini-2.5-flash to generate up to 20 Wikipedia page titles per query with relevance scores (1-5), including title matching with redirect name resolution.
  
  \item \textbf{lambdamart-rerank:} Hybrid first-stage retrieval (combining LLM, Pyterrier BM25, and BGE-M3 results) reranked with a trained LambdaMART model using dense/sparse scores, pageview counts, and PageRank as features.
  
  \item \textbf{gmn-rerank-500:} Hybrid first-stage retrieval (20 LLM results + top-500 BM25 + top-500 BGE-M3) reranked with Gemini-2.5-flash listwise reranking on all 1020 documents per query.
  
  \item \textbf{gm27q-LMART-1000:} Two-stage approach applying LambdaMART reranking to hybrid retrieval results (top-1000), followed by Gemma-27B-quantized reranking on top-500 documents.
  
  \item \textbf{gm27q-comb-500:} Limited hybrid retrieval (LLM results + top-200 BM25 + top-200 BGE-M3) reranked with Gemma-27B-quantized model.
\end{itemize}

\begin{table}[ht]
\centering
\caption{End-to-End System Results on Test Set}
\begin{tabular}{lccccc}
\toprule
System & R@10 & R@1000 & NDCG@1000 \\
\midrule
gmn-rerank-500 & 0.4341 & 0.6559 & 0.4106 \\
gm27q-LMART-1000 & 0.3617 & 0.6109 & 0.3339 \\
gm27q-comb-500 & 0.4196 & 0.5884 & 0.3848 \\
gemini-retrieval & 0.3457 & 0.3505 & 0.2962 \\
top\_model\_dense & 0.1720 & 0.5096  & 0.1620 \\
bge-m3 & 0.1447 & 0.5498 & 0.1492 \\
lambdamart-rerank & 0.1559 & 0.6109 & 0.1452 \\
\bottomrule
\end{tabular}
\label{tab:end_to_end_results}
\end{table}

Our best-performing system (\textit{gmn-rerank-500}) achieves a recall of 0.6559 and an NDCG of 0.4106. This demonstrates that combining multiple retrieval signals with strong LLM-based reranking provides substantial improvements over individual retrieval methods.

The top two systems, \textit{gmn-rerank-500} and \textit{gm27q-LMART-1000}, both employ hybrid retrieval followed by reranking but differ in their second-stage strategies. \textit{gmn-rerank-500} uses Gemini-2.5-flash reranking directly on the full hybrid results, while \textit{gm27q-LMART-1000} applies LambdaMART reranking first followed by Gemma-27B-quantized reranking on a smaller set. Although \textit{gmn-rerank-500} outperforms \textit{gm27q-LMART-1000} across all metrics, the latter offers a more computationally efficient approach yet still achieves competitive performance.

\section{Discussion}

\subsection{Network-based Ranking}

One of the hypotheses in our work is that network derived features would be viable for retrieval and reranking of documents by introducing bias toward popular or well-connected items in the knowledge graph.
These links are curated by editors themselves, and so we might expect that articles are linked together semantically through categorization.
Our graph construction takes these assumptions into account, where the first and second degree directed connections form the basis of a network-theoretic rank.
However, the number of connected components when doing so is far too small to be effective on a small query-set of items that we are not necessarily in control of, and would require constructing our own validation or test set so that these popular items are actual part of the recall set.
We increase connectivity of the graph across the full article-set by using the K-NN graph which uses a pre-computed document embedding, which ensures that every node has out-going links.
Because of the random-restart parameter of PageRank, information thus flows through all nodes of the network and relative rank is properly computed.

One of the simplest method for incorporating these features is by using it to rerank retrieved items. 
When we use network-theoretic rankings like out-degree, in-degree, or PageRank, we find that this is not an effective technique as the bias toward popular or network-importance is not necessarily the inductive bias needed to complete the tip-of-the tongue connection.
We experimented with this as part of a feature-based reranking system, and find that it does not contribute a high degree of information gain relative to the other features that we experimented with.

There were other ideas with the graph representation that we might try in the future: retrieval expansion via breadth-first search and seeding for synthetic dataset construction.
Retrieval expansion takes advantage of the fact that we capture neighborhood relations from the actual pagelink graph, and we can insert extra items into the set of retrieved item sorted by order of their importance in an attempt to increase recall on retrieval.
We could also use important network nodes to generate tip of the tongue query and result-sets by sampling articles proportional to their importance, and looking for similar articles within one or two hops of the sampled article as a target.

\subsection{Topic-Aware Dense Retrieval Performance}

The topic-aware multi-index dense retrieval (\textit{top\_model\_dense}) approach demonstrated a trade-off between computational efficiency and recall compared to the baseline BGE-M3 approach.
While partitioning the corpus into 24 topic-specific indexes and routing queries accordingly reduced search space and improved early precision metrics (NDCG@100: 0.1401 vs 0.122; recall@10: 0.172 vs 0.1447), it resulted in a slight decrease in overall recall@1000 (0.5096 vs 0.5498).
This suggests that topic-based partitioning may occasionally exclude relevant documents when query classification is imperfect or when target information spans multiple topical domains.
It is important to note that this submission represents first-stage retrieval only, without any reranking applied. As demonstrated by our other submissions that incorporated reranking stages, the addition of a reranker could significantly boost these results while still maintaining the computational advantages of topic-specific routing during the initial retrieval phase.

\subsection{LambdaMARAT Generalization}
We evaluated the impact of the LambdaMART reranker by comparing performance before and after using it to rerank the hybrid retrieval results on the test set. The hybrid results contained an average of 1,829 candidates per query. Following reranking, recall@1000 improved from 0.5700 to 0.6109, which shows that the model effectively surfaces relevant documents into the top 1,000.
However, the Reciprocal Rank fell significantly from 0.2838 to 0.0601. This indicates that while LambdaMART excels at identifying relevant documents, it often displaces those originally at the top to lower positions. This explains the low NDCG observed for the \texttt{lambdamart-rerank} system. This result is different from what we observed in the training and development sets, where the precision of the result does not decrease. This shows that the model does not generalize well on the test set. However, we can still use LamdbaMART as a high-recall filter in this pipeline and the subsequent application of LLM reranking helps to recover the precision of the final result.

\subsection{Performance Analysis on Data Source}
The test dataset comprises queries from three sources \cite{trec-tot2025-overview}: (1) human-elicited queries from NIST assessors; (2) synthetic queries generated from GPT-4o and Llama-3.1-8B-Instruct; and (3) the MS-ToT Dataset. The performance of our best system (\textit{gmn-rerank-500}) varies significantly across these subsets. The NIST assessor queries achieved a recall@1000 of 0.89 and NDCG@1000 of 0.65. In contrast, the MS-ToT and synthetic datasets proved more challenging, with recall@1000 reaching 0.59 and 0.57, and NDCG@1000 at 0.32 and 0.37, respectively. These results align with the track coordinators' observations regarding the lower relative difficulty of the NIST-sourced queries.

\section{Conclusions}

This work addresses the TREC Tip-of-the-Tongue task through a comprehensive two-stage retrieval pipeline that strategically combines multiple retrieval methods with both learned and LLM-based reranking approaches.

Our work includes: (1) building a hybrid first-stage retrieval strategy that effectively merges results from LLM, sparse, and dense retrieval methods, (2) developing a topic-aware multi-index dense retrieval approach that improves computational efficiency while maintaining competitive recall through semantic document partitioning, and (3) training a learned LambdaMART reranker on both official and synthetic data with features from retrieval scores and document popularity metrics.

Our best system achieves substantial improvements over baseline methods by merging hybrid retrieval with strong LLM-based reranking. The results demonstrate that leveraging multiple retrieval methods is crucial for capturing complementary aspects of relevance. While LLM-based reranking achieves superior performance, using a learned LambdaMART reranker with a weaker LLM provides a computationally efficient alternative with competitive results.

\section*{Acknowledgements}

We thank the Data Science at Georgia Tech Applied Research Competition group (DS@GT ARC) for their support.
This research was supported in part through research cyberinfrastructure resources and services provided by the Partnership for an Advanced Computing Environment (PACE) at the Georgia Institute of Technology, Atlanta, Georgia, USA \cite{PACE}.

\bibliographystyle{ACM-Reference-Format}
\bibliography{tot-bib}

\appendix

\section{LambdaMART Hyperparameter Tuning Details}
\label{sec:lambdamart_tuning}

We explored hyperparameter tuning using both grid search and random search strategies. The parameter ranges for each search method are shown in Table~\ref{tab:lambda_hyperparameter_experiment}. Both approaches trained models with pairwise loss and NDCG as the evaluation metric, with each model trained over 100 rounds and 20 trials performed on the random search.

\begin{table}[ht]
    \centering
    \caption{LambdaMART hyperparameter search space.}
    \begin{tabular}{lcc}
        \toprule
        Hyperparameter & Grid search & Random search \\
        \midrule
        Learning Rate & [0.05, 0.1, 0.2] & [0.01, 0.05, 0.1, 0.2] \\
        Max Tree Depth & [4, 6, 8] & [3, 4, 6, 8, 10] \\
        Min Child Weight & [1, 5] & [1, 3, 5, 7] \\
        Data Sampling Ratio & [0.8, 1.0] & [0.6, 0.8, 1.0] \\
        Feature Sampling Ratio & [0.8, 1.0] & [0.6, 0.8, 1.0] \\
        Split Loss Threshold & 0 & [0, 0.1, 0.2, 0.5] \\
        \bottomrule
    \end{tabular}
    \label{tab:lambda_hyperparameter_experiment}
\end{table}

\subsection{Model Variants: v5 vs v6}

The hyperparameter search generated two model variants: v5 from grid search and v6 from random search. Their final hyperparameters and training metrics are shown in Table~\ref{tab:lambda_final_hyperparameter}.

\begin{table}[ht]
    \centering
    \caption{LambdaMART final hyperparameters and metrics for v5 and v6.}
    \begin{tabular}{lcc}
        \toprule
          & v5 (grid) & v6 (random) \\
        \midrule
        Learning Rate & 0.2 & 0.2 \\
        Max Tree Depth & 6 & 8  \\
        Min Child Weight & 5 & 1 \\
        Data Sampling Ratio & 1.0 & 0.8 \\
        Feature Sampling Ratio & 1.0 & 0.8\\
        Split Loss Threshold & 0 & 0.5 \\
        Train NDCG & 0.892 & 0.917 \\
        Validation NDCG & 0.869 & 0.867 \\
        \bottomrule
    \end{tabular}
    \label{tab:lambda_final_hyperparameter}
\end{table}

While v6 achieved higher training NDCG (0.917 vs 0.892), it did not generalize as well to unseen dev and synthetic datasets. The validation NDCG scores were similar (0.869 vs 0.867), but v6's more aggressive hyperparameters (deeper trees, lower sampling ratios) led to overfitting. Additionally, the simpler parameter configuration of v5 (higher sampling ratios, shallower trees) makes it more robust and interpretable. Therefore, v5 was selected as the final model for all main results.

\begin{figure}[ht]
    \centering
    \includegraphics[width=0.9\linewidth]{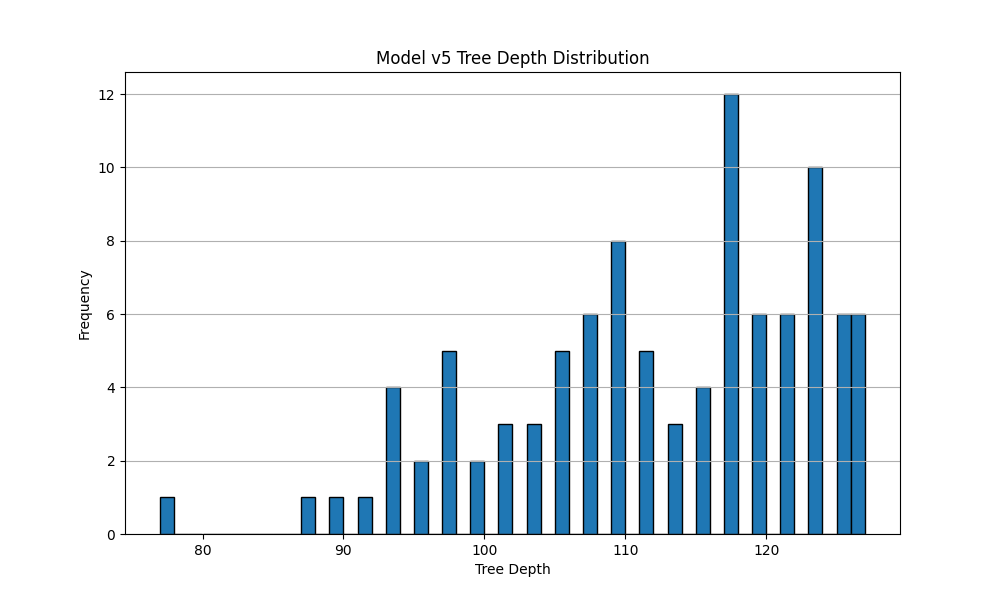}
    \hfill
    \includegraphics[width=0.9\linewidth]{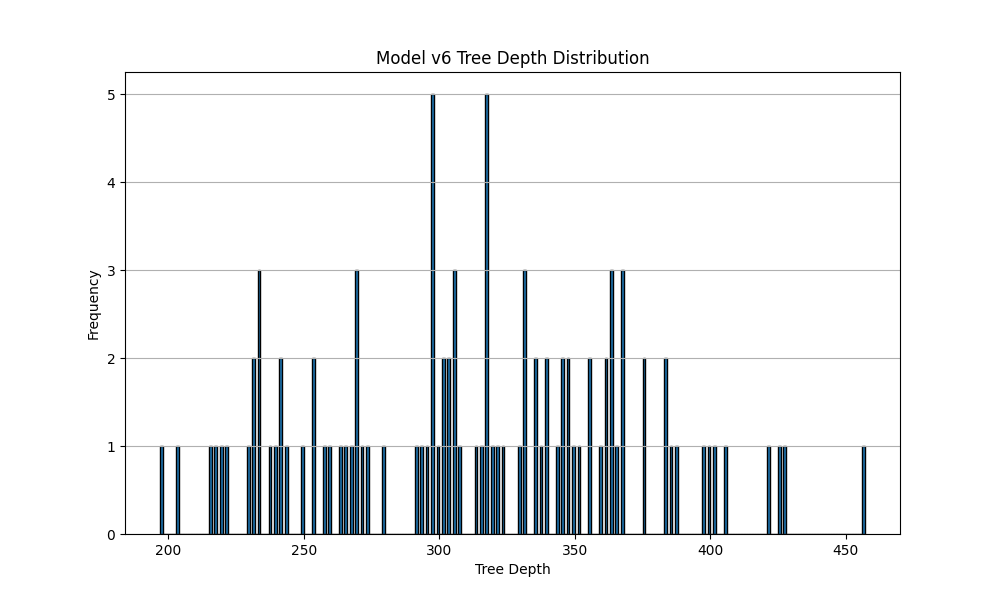} 
    \caption{Tree depth distribution for LambdaMART v5 and v6 models.}
    \label{fig:lambda_model_visualizations}
\end{figure}

Figure~\ref{fig:lambda_model_visualizations} shows the tree depth distribution comparison between v5 and v6.

\onecolumn

\section{Prompt for Open Domain Query Generation}
\label{sec:prompt}

\fbox{
\begin{minipage}{0.8\textwidth}
Let's do a role play. You are now a person who vaguely remembers something called \{ToTObject\}. You are trying to recall the name by posting a verbose post in an online forum like Reddit describing it. Generate a post of around 200 words about \{ToTObject\}. Your post must describe a vague memory without revealing its exact name. People on the forum must have a hard time figuring out what you are looking for. The answer should be difficult to find in search engines, so avoid using obvious keywords. I will provide you with some basic information, and you must follow the guidelines to create a post.
\newline\newline
Information about \{ToTObject\}:

\{Psg\}

Guidelines:
MUST FOLLOW:
1. Reflect the imperfect nature of memory with phrases that express doubt or mixed recollections, avoiding direct phrases like "I'm not sure if it is true, but". 
2. Do not directly specify the name of what you are trying to recall.\newline
3. Refer to it in an ambiguous way using descriptions instead of names.\newline
4. Maintain a casual and conversational tone throughout the post, making sure it sounds natural and engaging without using formal structures.\newline
5. Provide vivid but ambiguous details to stir the reader's imagination while keeping them guessing.\newline
6. Use the provided information only as inspiration to craft a unique and engaging narrative, avoiding any direct replication of the given phrases.\newline
7. Start directly with your post, avoiding formal greetings like "Hello" or "Hey everyone."\newline
8. Start directly with your post, without describing your state of mind like "So, there's this", "I remember", "I've been thinking about".

COULD FOLLOW:\newline
1. Share a personal anecdote related to what you are trying to recall, but avoid common phrases like "When I was young." Instead, find unique ways to set the scene.\newline
2. Focus on sensory details like the overall mood, sounds, and emotional impact of the memory, using vivid descriptions.\newline
3. Draw comparisons with other similar things in a nuanced way that doesn't directly echo well-known examples.\newline
4. Introduce a few incorrect or mixed-up details to make the recollection seem more realistic and harder to pinpoint.\newline
5. Describe particular scenes or moments using ambiguous terms or partial descriptions.\newline
6. End the post by encouraging responses with genuine, open-ended questions for help.\newline
Generate a post based on these guidelines. Wrap the post in a code block.
\end{minipage}
}

\section{Prompt for Zero-shot TOT Retrieval}
\label{sec:prompt_retrieval}

\fbox{
\begin{minipage}{0.8\textwidth}
Identify up to 20 entities that are titles of Wikipedia pages which answer the following tip-of-the-tongue query.\newline\newline
For each entity, provide the Wikipedia page title and a relevance score from 1-5, where:\newline\newline
- 1 = Irrelevant to the query\newline
- 2 = Somewhat relevant\newline
- 3 = Moderately relevant\newline
- 4 = Highly relevant\newline
- 5 = Most relevant and directly answers the query\newline\newline
Return a JSON object containing the entity titles and their relevance scores.\newline\newline
TOT Query: \{query\}

\end{minipage}
}

\twocolumn

\end{document}